\documentclass{an-think}
\usepackage{graphicx,fancyhdr}
\usepackage{times}

\renewcommand{\vec}[1]{\mbox{\boldmath $#1$}}

\begin{document}

\sloppy
\newcommand{\kms}{km\,s$^{-1}$}
\newcommand{\Halpha}{H$\alpha$}

\title{Anti-solar differential rotation}

\author{L.~L.~Kitchatinov\inst{1,}\inst{2}\fnmsep\thanks{lkitchatinov@aip.de,~
kit@iszf.irk.ru},
G.~R\"udiger\inst{1}\fnmsep\thanks{gruediger@aip.de}\\}

\institute{Astrophysical Institute Potsdam (AIP), An
der Sternwarte 16, D-14482 Potsdam, Germany
\and
Institute for Solar-Terrestrial Physics, P.O.~Box 4026, Irkutsk, 664033,
Russian Federation
}

\date{Received; accepted; published online}

\abstract{The differential rotation of anti-solar type detected by
observations for several stars may result from a fast meridional
flow. The sufficiently intensive meridional circulation may be
caused by large-scale thermal inhomogeneities or, perhaps, by
tidal forcing from a companion star. First results of simulations
of the anti-solar rotation of a giant star with magnetically
induced thermal inhomogeneities are presented. Perspectives for
observational check of the theory are discussed.
 }
\maketitle
\section{Introduction}
This paper discusses possible theoretical explanations for observations
of non-uniform stellar rotation with the angular velocity increasing with
latitude. It also suggests the possibilities for observational check of
the theory.

The anti-solar rotation is relatively seldom to observe. The
solar-like case of  rotation rate decreasing with latitude is more
frequent to detect (Strassmeier \cite{S03}, Petit et al.
\cite{PDC04}). The relatively fast rotation of the equator
represents also the case which theory succeeded to explain. Theory
of angular momentum transport by rotating turbulence and numerical
simulations both show meridional fluxes of the angular momenrtum
towards equator (R\"udiger \& Hollerbach \cite{RH04}). Models for
differential rotation based on that theory provide rotation law
for the Sun (Kitchatinov \& R\"udiger \cite{KR95}, Kitchatinov
\cite{K04}) in close agreement with  helioseismology (Shou et al.
\cite{Sea98}). Observations of solar-type rotation for AB~Dor
(Donati \& Collier Cameron \cite{DC97}) and LQ~Hya (Kov\'ari et
al. \cite{Kea04}) can be reproduced with reasonable accuracy
(Collier Cameron et al. \cite{CBKD01}, Kitchatinov et al.
\cite{KJD00}), and theoretical trend for differential rotation of
main-sequence dwarfs to decrease with spectral type (Kitchatinov
\& R\"udiger \cite{KR99}) was given some observational support
(Collier Cameron \cite{CBKD01}, Petit et al. \cite{PDC04}).

However, observations of anti-solar rotation are already numerous
enough to demand a revision of theory. The standard hydrodynamical
models cannot reproduce this case. The most plausible way for such
a revision is to include some additional driver of meridional flow.
We shall see in the next section that the fast meridional circulation
is the most clear theoretical possibility for producing differential
rotation of the anti-solar type. Observations may help a lot in
understanding the nature of stellar differential rotation by checking
whether all the stars which show anti-solar rotation do indeed possess
a fast meridional flow.

The fast meridional motion can result from a barocline driving due
to large thermal spots or from tidal forcing by a close companion.
The guess is suggested by both theoretical arguments and
statistics of differential rotation observations for individual
stars summarized by Strassmeier (\cite{S03}). All stars with
detected anti-solar rotation belong to one of two (yet small)
groups: (i) close binaries, or (ii) rapidly rotating giants with
large-scale thermal inhomogeneties (Hackman et al. \cite{HJT01},
Strassmeier et al. \cite{SKW03}). The suggestion that all
anti-solar rotators are spotted giants or close binaries is more
speculative and less certain than the conclusion on the role of
meridional flow but may also be a subject for observational
verification.

After a general discussion of possible origin of anti-solar
rotation in the next section, section~\ref{model} outlines a
numerical model for equatorial deceleration on a giant star with
large-scale thermal inhomogeneties caused by magnetic fields.
Results of first simulations are presented and discussed in
section~\ref{results}. The final section~\ref{to_observe} suggests
the ways for observational verification of the theory.
\section{General consideration}\label{general}
\subsection{Anti-solar rotation for fast meridional flow}
Possible reason for anti-solar differential rotation can be
inferred from consideration of a global axisymmetric flow in a
spherical convective shell. The flow velocity, $\bf u$, can be
expressed in terms of the stream-function, $\psi$, of the
meridional circulation and angular velocity, $\Omega$, using the
standard spherical coordinates, $r,\vartheta , \varphi$,
\begin{equation}
 {\vec u} = \left({1\over\rho r^2\sin\vartheta}{\partial \psi\over\partial
 \vartheta} ,\ {-1\over\rho r\sin\vartheta}{\partial \psi\over\partial r},\
 r \sin\vartheta\ \Omega \right).
 \label{1}
\end{equation}
The flow obeys the steady motion equation,
\begin{eqnarray}
   \rho\left({\vec u}\cdot\nabla\right){\vec u}
   &+& \left( \left({\vec B}\cdot\nabla\right){\vec B}
   - \nabla B^2/2\right)/\left( 4\pi\right)
   \nonumber \\
    &+& \nabla p - \rho\nabla\Phi = -
   {\rm div}\left(\rho Q\right) .
   \label{2}
\end{eqnarray}
where $\vec B$ is magnetic field, $\Phi$ is gravity potential, and
$\hat Q$ is the correlation tensor of fluctuating velocities, $\vec u'$,
\begin{equation}
 Q_{ij} = \langle u'_iu'_j\rangle .
 \label{3}
\end{equation}

The most promising for producing anti-solar rotation is a fast
meridional flow. One can find that from the equation for angular
velocity which results as $\phi$-component of Eq.~(\ref{2}),
\begin{equation}
  {1\over\sin^2\vartheta}\ {\partial\psi\over\partial r}\
  {\partial\left(\sin^2\vartheta\Omega\right)\over\partial\vartheta} -
  {1\over r^2}\ {\partial\psi\over\partial\vartheta}\
  {\partial\left( r^2\Omega\right) \over\partial r}\ \  ... = 0 ,
\label{4}
\end{equation}
where only the contribution of meridional flow is written
explicitly. The dotted terms in (\ref{4}) can be neglected if the
flow is fast enough. Then, the solution can be found,
\begin{equation}
 \Omega (r,\vartheta ) = {F(\psi )\over r^2\sin^2\vartheta} ,
 \label{5}
\end{equation}
which describes almost certainly the inhomogeneous rotation of the
anti-solar type. The angular momentum is conserved along stream
lines of a fast meridional flow (R\"udiger \cite{R89}). $F$  in
equation (\ref{5}) is an arbitrary function. Additional conditions
are required to define it. Whatever the conditions could be, the
solution (\ref{5}) tends to describe the anti-solar rotation.
Indeed, the stream-function, $\psi$, is constant along the stream
lines. Therefore, the angular velocity (\ref{5}) varies along the
stream lines in such a way that it increases when the lines
approach the rotation axis at high latitudes. The conclusion does
not depend on whether the flow is poleward or equatorward (on the
surface), in any case the fast flow tends to produce the
anti-solar differential rotation.
\subsection{Barocline flow}
A barocline meridional flow can be sufficiently fast for
supporting the anti-solar rotation. The barocline term
contributes the equation for a steady meridional flow (Kitchatinov
\& R\"udiger \cite{KR99}),
\begin{equation}
  {\cal D}\left(\psi\right) =
  {1\over\rho^2}\left(\nabla\rho\times\nabla p\right)_\varphi\ ...\ ,
  \label{6}
\end{equation}
when the surfaces of constant density and pressure do not
coincide. In equation (\ref{6}), the left part accounts for
resistance to meridional circulation by eddy viscosity and only
the barocline source is written explicitly in the right. It is
convenient for our purposes to express it in terms of specific
entropy, $S$, and gravity potential, $\Phi$, as it was done by
Kitchatinov \& R\"udiger (\cite{KR99}),
\begin{equation}
  {\cal D}\left(\psi\right) =
  {1\over c_{\rm p}r}\left( {\partial S\over\partial r}
  {\partial\Phi\over\partial\vartheta} -
  {\partial S\over\partial\vartheta}
  {\partial\Phi\over\partial r}
  \right)\ ...\ .
  \label{7}
\end{equation}
The entropy and gravity distributions are normally close to
spherical symmetry and the resulting barocline flow is small. This
is why the advection dominated states did not emerge and
anti-solar rotation was not found in former simulations.

Meridional circulation can be fast, however, if considerable
deviations from spherical symmetry are available in gravity or
temperature distributions. The asymmetric gravity is present in
binary systems and asymmetric temperature is typical of giants
with their large thermal spots. Anti-solar rotation can be
expected for these cases. The guess agree with statistics of
anti-solar rotation detections summarized by Strassmeier
(\cite{S03}). Between nine anti-solar rotators, six belong to
close binaries and two - to giant stars with large thermal
inhomogeneities on their surfaces, the remainder star is LQ~Hya
for which different observations disagree on the sense of its
differential rotation.

Later on, we focus on the case of thermal asymmetry to treat it in
a more quantitative way.
\subsection{Convective heat transport in magnetized fluids}
Stellar spots are believed to be magnetic by origin. To account
for  the magnetic field influence on thermodynamics, we include
magnetic quenching of convective heat flux,
\begin{equation}
  F^{\rm conv}_i = -\rho T\chi_{ij}{\partial S\over\partial r_j},\ \ \ \
  \chi_{ij} = \varphi\left(\beta\right)\chi^0_{ij},
  \label{8}
\end{equation}
where $\chi^0$ is the eddy thermal conductivity for nonmagnetic
case  and $\varphi$ is the quenching function of the normalized
field strength $\beta = B/\sqrt{4\pi\rho\langle{u'}^2\rangle}$.
The function $\varphi$ is steadily decreasing with $\beta$. The
quenching function was derived in the paper by Kitchatinov et al.
(\cite{KPR94}) where it is given as $\varphi_\chi$-function.
\section{The model}\label{model}
Our present model is very close to its previous version
(Kitchatinov \& R\"udiger 1999). We describe the model only
briefly focusing on where it is different from the former
formulation.

The main difference is that we include now axisymmetric magnetic
field governed by the steady induction equation,
\begin{equation}
   \nabla\times\left({\vec u}\times{\vec B} -
   \eta_{_{\rm T}}\left(\beta\right)\nabla\times{\vec B}\right) = 0.
   \label{9}
\end{equation}
Eddy magnetic diffusivity is defined in terms of superadiabaticity
of the stratification and includes quenching by magnetic field and
rotation,
\begin{equation}
  \eta_{_{\rm T}} = -{\tau\ell^2 g\over 12 c_p}{\partial S\over\partial r}
  \phi(\Omega^* ) \varphi (\beta ) ,
\label{10}
\end{equation}
where $\ell$ and $\tau$ are mixing length and time respectively,
the  function $\phi$ is given in Kitchatinov et al.
(\cite{KPR94}), the magnetic quenching function, $\varphi$ is the
same as before, and $\Omega^*$ is the Coriolis number,
\begin{equation}
  \Omega^* = 2\tau\Omega .
\label{11}
\end{equation}
Anisotropy of magnetic diffusion is neglected in (\ref{9}). We
neglect  also the anisotropy of eddy viscosity, but keep the
rotationally induced anisotropy of thermal conductivity because
the differential rotation models cannot function normally without
it (R\"udiger et al. \cite{REKK04}). Magnetic quenching of all the
turbulent transport coefficients including the $\Lambda$-effect
(R\"udiger \cite{R89}) was described by same quenching function
$\varphi (\beta )$ (\ref{8}). The axisymmetric magnetic field can
be expressed in terms of the toroidal field $B$ and potential $A$
of the poloidal field:
\begin{equation}
  {\vec B} = \left( {1\over r^2\sin\vartheta}{\partial A\over\partial\vartheta} ,
  {-1\over r\sin\vartheta}{\partial A\over\partial r} , B \right).
\label{12}
\end{equation}

Our numerical model solved a system of five joint equations for
angular velocity, meridional flow, entropy, poloidal and toroidal
components of the  magnetic field.

The induction equation (\ref{9}) does not include the
$\alpha$-effect  (Krause \& R\"adler \cite{KR80}) of turbulent
dynamo. Therefore, our model cannot support any dynamo. The
magnetic field was involved through the boundary condition of a
steady radial field penetrating the convection zone at the inner
boundary, $r_i$, from the radiative core. The bottom field was
prescribed by the steady potential $A$ at the inner (bottom)
boundary,
\begin{equation}
  A = r_i^2 B_0\left( 1 - \cos^{2n}\vartheta\right),\ \ \ \
  B_0 = {\Phi^{\rm m}\over 2\pi r_i^2} ,
\label{13}
\end{equation}
where $\Phi^{\rm m}$ is the magnetic flux of the dipolar field per
hemisphere, $n$ is the parameter controlling the latitudinal
distribution of the field, the larger is $n$ there more
concentrated to poles is the poloidal field. Computations for
various $n$ were performed. Unless otherwise stated, the results
of the next section correspond to $n=3$. The condition (\ref{13})
for the inner boundary can be understood as penetration of a relic
field stored in the radiative core into the convection zone. In
convection zone proper, the field is subject to turbulent
diffusion and advection, so that toroidal field is produced by the
differential rotation.

The other boundary conditions were zero radial velocity and zero
stress at the top and bottom, constant heat flux at the bottom and
black-body radiation at the top, superconducting condition at the
bottom for the magnetic field and vacuum condition on the top.
\section{Results and discussion}\label{results}
We performed our simulations for a giant star evolved from the
main-sequence. The parameters of the star were taken form an
appropriate model for stellar structure (Herwig et al.
\cite{HBSE97}), some of them are given in the table. The assumed
rotation rate is marginal for Doppler imaging. The rate is,
however, several times larger than normal for this type of stars
(Gray \cite{G89}).

\begin{table}\label{t1}
\caption{Parameters of the star}
\begin{tabular}{lllll}
  \hline
  $M/M_\odot$ & $R/R_\odot$ & $L/L_\odot$ & $r_i/R$ & $V_{\rm eq}$, km/s \\
  \hline
  2.5 & 7.91 & 42.1 & 0.61 & 15 \\
  \hline
\end{tabular}
\end{table}

\begin{figure}[ht]
   \centering
   \includegraphics[width=8.5cm]{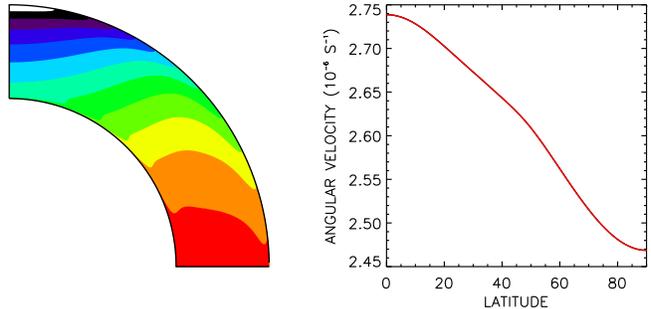}
   \caption{Differential rotation for nonmagnetic case is of
            the solar type. Angular velocity increases from
        poles to equator by about 10\%.}
   \label{f1}
\end{figure}

The differential rotation simulated for the nonmagnetic state is
shown in Fig.~\ref{f1}. The standard equatorial acceleration of
about 10\% was found for this case.

\begin{figure}[ht]
   \centering{
   \includegraphics[width=4.2cm]{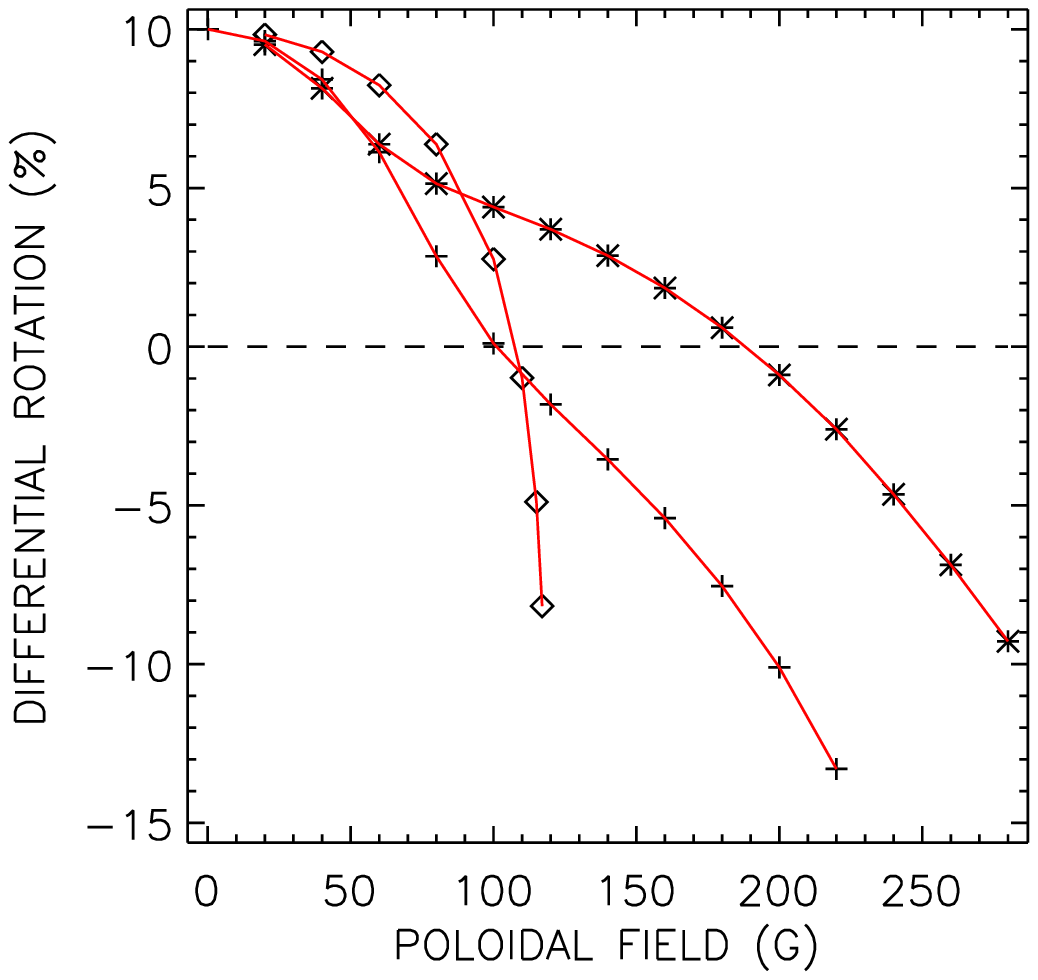}
   \includegraphics[width=4.2cm]{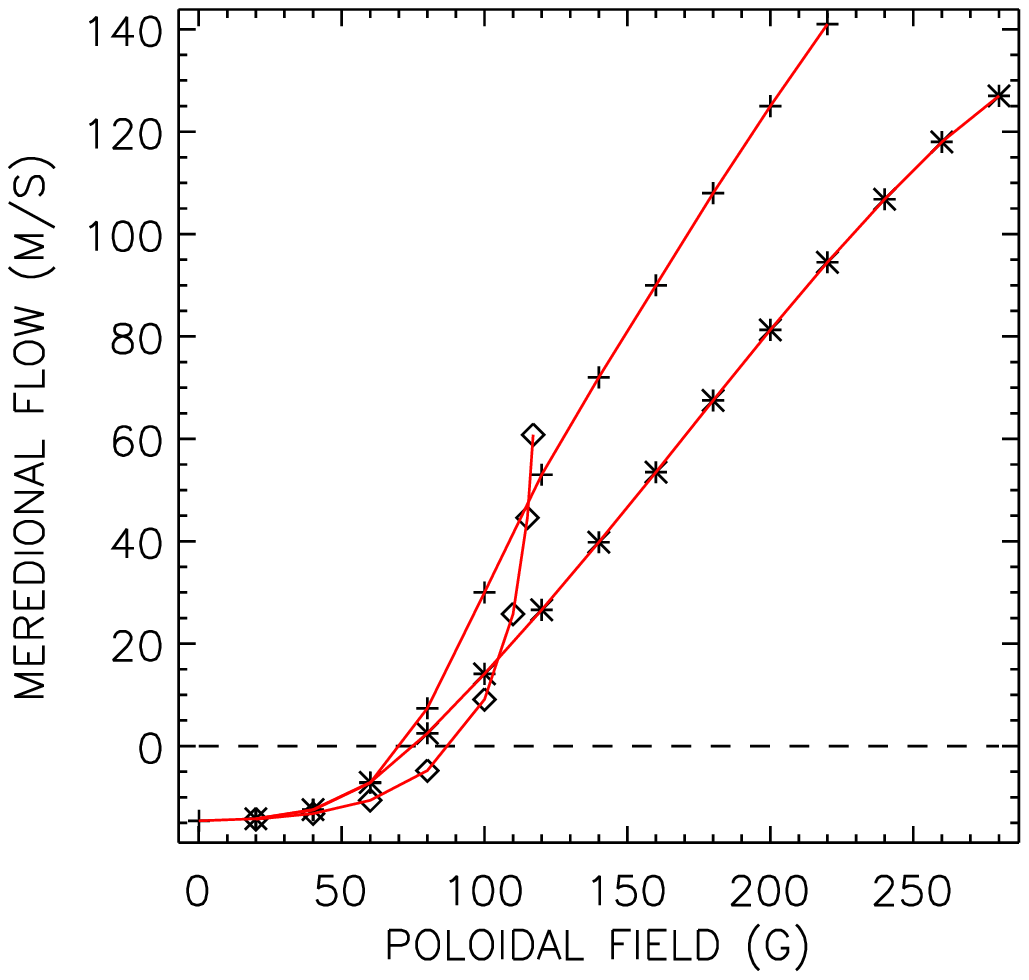}}
   \caption{Left panel: surface differential rotation,
      $100\left(\Omega^{\rm eq}-\Omega^{\rm pole}\right)/\Omega_0$,
      as function of the poloidal field amplitude,
      $B_0$ (\ref{13}). Actually computed points are marked by
      signs. Different signs correspond to different values
      of the $n$-parameter of the bottom profile (\ref{13})
      of the poloidal field: $n=1$ (diamonds), $n=3$
      (crosses), $n=5$ (stars). Right panel: Surface
      meridional flow at 45$^\circ$-latitude. Positive
      values mean poleward flow.}
   \label{f2}
\end{figure}

Figure~\ref{f2} shows how the global surface flow varies with the
amplitude of prescribed poloidal field (\ref{13}) for several
latitudinal profiles of the field. As the field increases, the
meridional flow reverses to poleward orientation and then grows
steadily. When the flow becomes sufficiently fast, the
differential rotation reverses to anti-solar case. We were able to
follow the dependencies up to the value of order $B_0 \sim 100 ...
300$~G depending on the bottom profile (\ref{13}) of the poloidal
field. Our numerical code based on the relaxation method did not
converge for stronger fields. The numerical instability, probably,
signals on the onset of the physical flux-concentration
instability (Kitchatinov \& Mazur \cite{KM00}) via which the cool
magnetic spots are formed. By this reason, we did not have true
magnetic spots but smooth magnetically induced thermal
inhomogeneities in our simulations.

\begin{figure}[ht]
   \centering
   \includegraphics[width=8.5cm]{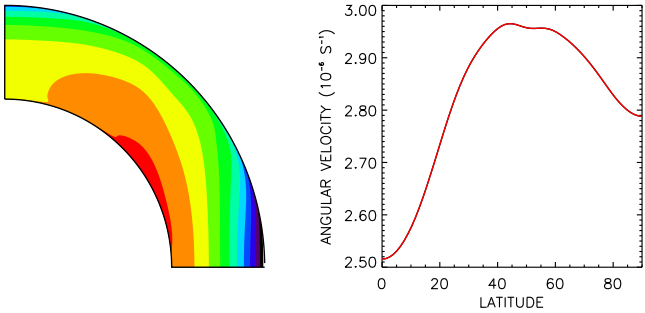}
   \caption{Simulated differential rotation for poloidal
            field amplitude $B_0=200$~G. Angular
            velocity {\sl increases} from equator to
            pole by about 10\%. Low and high
            latitudes have opposite sense of the
            latitudinal rotation
            inhomogeneity.}
   \label{f3}
\end{figure}

\begin{figure}[ht]
   \centering
   \includegraphics[width=8.5cm]{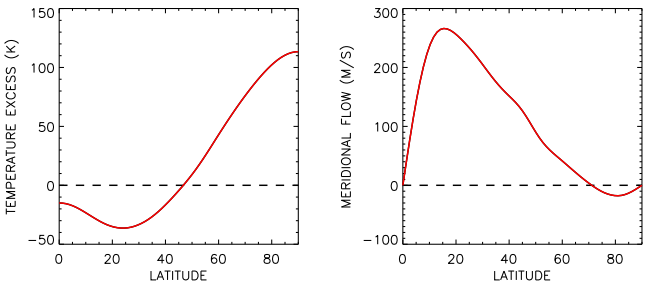}
   \caption{Left panel: Deviation of surface temperature
            from its mean value as a function of latitude
            for same model as Fig.~\ref{f3}. Right panel:
            surface meridional flow. $B_0 = 200$~G.}
   \label{f4}
\end{figure}

Figure~\ref{f3} illustrates typical case of anti-solar rotation of
present  simulations. Angular velocity increases from equator to
pole but not steadily. In agreement with above qualitative
arguments, the increase occurs in the same latitude range where
relatively fast poleward meridional flow can be observed in
Fig.~\ref{f4}. The surface rotation of Fig.~\ref{f3} cannot be
approximated by traditional $\cos^2\vartheta$ profile. It may be
reasonable to use higher-order terms in approximations of the
observed anti-solar rotation laws. Expansion in terms of
Gegenbauer polynomials (R\"udiger \cite{R89}),
\begin{equation}
  \Omega\left(\vartheta\right) = \sum_{n=1}^N {\omega_n\over\sin\vartheta}
   P_{2n-1}^1\left(\cos\vartheta\right) ,
\label{14}
\end{equation}
may be convenient, especially when rotation law variations with time are concerned.

The meridional flow of Fig.~\ref{f4} is driven by the barocline
forcing due to the thermal depletion which occupies same region of
latitudes. As was already mentioned, there were no true thermal
spots in the simulations. Nevertheless, the thermal depletion of
Fig.~\ref{f4} was produced by magnetic quenching of convective
heat flux of Eq.~(\ref{8}).

\begin{figure}[ht]
   \centering
   \includegraphics[width=8.5cm]{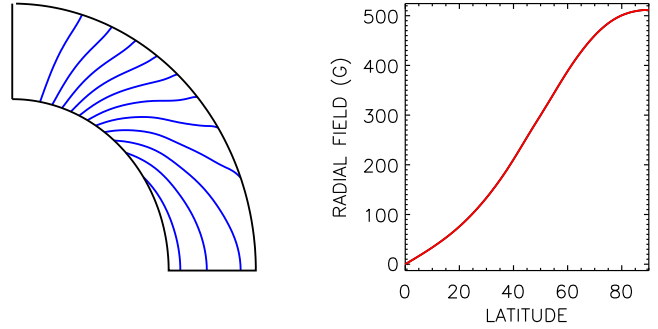}
   \caption{Left panel: lines of the poloidal magnetic
            field for same model as Fig.~\ref{f3}.
            Right panel: surface field strength.}
   \label{f5}
\end{figure}

\begin{figure}[ht]
   \centering
   \includegraphics[width=8.5cm]{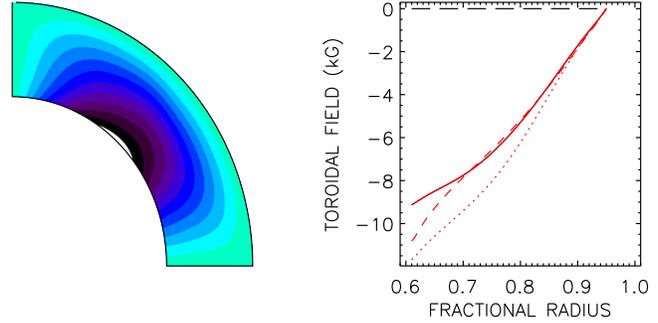}
   \caption{Toroidal magnetic field for same model as
            Fig.~\ref{f3}.
            Right panel shows radial profiles of the
            field for latitudes of 20$^\circ$ (full line),
            40$^\circ$ (dotted),
            and 60$^\circ$ (dashed). }
   \label{f6}
\end{figure}

The magnetic field structure and amplitude are illustrated by
Figs.~\ref{f5} and \ref{f6}. The poloidal field amplitude, $B_0 =
200$~G, is the mean value. Its maximum strength in the polar
region is about 2.5 times larger. It is not poloidal field,
however, which makes the principal influence on convective heat
flux (\ref{8}) and produces the temperature depletion of
Fig.~\ref{f4}. The toroidal field of Fig.~\ref{f6} is stronger by
about one order of magnitude.

The anti-solar rotation of Fig.~\ref{f3} can be interpreted along
the following sequence. Toroidal magnetic field of Fig.~\ref{f6}
produces thermal depletion of Fig.~\ref{f4} by suppression of the
convective heat flux (\ref{8}). The thermal depletion drives a
relatively fast meridional flow of Fig.~\ref{f4} via the barocline
forcing of Eq.~(\ref{7}). The meridional flow tends to make the
angular momentum uniform along its stream-lines in accord with the
solution (\ref{5}) of the angular velocity equation (\ref{4}) for
the advection-dominated case thus producing relatively slow
rotation of the equator. It should be noted, however, that
convective fluxes of angular momentum (the $\Lambda$-effect,
R\"udiger \& Hollerbach \cite{RH04}) does also take part in
formation of the differential rotation, and we return to the
beginning of our interpretation by noticing that the toroidal
magnetic field is produced from poloidal one by the differential
rotation.

The simulated anti-solar rotation of preceding section is the
first theoretical indication for this type of differential
rotation. Nevertheless, it is tempting to notice points of yet
qualitative agreement with observations of the giant star HD~31993
by Strassmeier et al. (\cite{SKW03}). The star shows differential
rotation of about 12\% of anti-solar type. The dark spots were
observed on low latitudes and they had temperature contrast of
about 200$^\circ$~K only which is compatible with
(latitude-averaged) moderate temperature depletion of
Fig.~\ref{f4}. In a further agreement, the observed star had a
warm pole. It is not clear, however, whether meridional flow was
as fast as required by our theoretical model to produce anti-solar
rotation.

Unfortunately, HD~31993 remains the only giant star for which the
anti-solar type of rotation was detected incontroversially. The
other candidate, HD~199178, is under debates (Hackmann et al.
\cite{HJT01}, Petit et al. 2004).

The anti-solar rotation was detected for six close binaries also
(Strassmeier \cite{S03}). Consideration of section~\ref{general}
suggests this type of star as another possibility for equatorial
deceleration though no quantitative model for binaries was
developed so far.
\section{What to observe?}\label{to_observe}
It seems to be a quite general theoretical statement that
relatively fast meridional flow should be present whenever
differential rotation is of anti-solar type. Accordingly, it is
fundamentally important to know whether observations can confirm
that all stars showing relatively slow equatorial rotation do
simultaneously possess a large meridional flow.

It is appropriate to estimate how fast the meridional circulation
should be for producing the anti-solar differential rotation.
Significance of meridional flow is defined by the Reynolds number,
${\rm Re} = u^{\rm m} R/\nu_{_{\rm T}} \simeq u^{\rm
m}R\tau/\ell^2$ ($\ell$ and $\tau$ are mixing length and time
respectively). Judging from Fig.~\ref{f2}, the \emph{poleward}
meridional flow of
\begin{equation}
 u^{\rm m} \geq 30{\ell^2\over\tau R}
 \label{mf}
\end{equation}
should be sufficient to support the equatorial deceleration. The
fast circulation is expected to drive differential rotation of
anti-solar type independently of whether the direction of the
surface flow is towards pole or equator. Estimations by R\"udiger
(\cite{R89}) suggest, however, that \emph{equatorward} flow should
be several times faster compared to Eq.~(\ref{mf}) to produce
anti-solar rotation.

It may be noticed that the meridional circulation has been
recognized as an important ingredient of stellar dynamos
(Choudhuri et al. \cite{CSD95}, Dikpati \& Gilman \cite{DG01},
Bonnano et al. \cite{BERB02}). Accordingly, observational
detections of the meridional flow would be helpful for stellar
dynamo theory also.

Another possibility for observational verification of the above
theoretical predictions is to check whether all  anti-solar
rotators do indeed belong to one of two groups: spotted giants or
close binaries. The suggestion that a fast meridional flow can be
found only for the two groups of stars is, however, more
speculative and less certain than the necessity of the meridional
flow itself for rotation laws of anti-solar type.

Measuring differential rotation and meridional flow on giant stars
can be a difficult observational task because of relatively long
rotation periods of giants. The task may be a natural subject for
robotic astronomy because long observational series may be
required.
\acknowledgements L.L.K. is grateful to A.I.P. for its hospitality
and visitors support.


\begin{thebibliography}{}
\bibitem[2002]{BERB02}
   Bonnano~A., Elstner~D., R\"udiger~G., Belvedere~G.: 2002, A\&A 390, 673
\bibitem[1995]{CSD95}
   Choudhuri~A.~R., Sch\"ussler~M., Dikpati~M.: 1995, A\&A 303, L29
\bibitem[2001]{CBKD01}
   Collier Cameron~A., Barnes~J.~R., Kitchatinov~L., Donati~J.-F.: 2001,
   Differential Rotation in Young Low-mass Stars. In ASP Conf. Ser. 223:
   11th Cambridge Workshop on Cool Stars, Stellar Systems and the Sun, p.251
\bibitem[2001]{DG01}
   Dikpati~M., Gilman~P.: 2001, ApJ 559, 428
\bibitem[1997]{DC97}
   Donati~J.-F., Collier Cameron   ~A.: 1997, MNRAS 291, 1
\bibitem[1989]{G89}
   Gray~D.~F.: 1989, ApJ 347, 221
\bibitem[2001]{HJT01}
   Hackman~T., Jetsu~L., Tuominen~I.: 2001, A\&A 374, 171
\bibitem[1997]{HBSE97}
   Herwig~F., Bl\"ocker~T., Sch\"onberner~D., El~Eid~M.: 1997, A\&A 324, L81
\bibitem[2004]{K04}
   Kitchatinov~L.~L.: 2004, Astron. Rep. 48, 153
\bibitem[2000]{KM00}
   Kitchatinov~L.~L., Mazur~M.~V.: 2000, Sol. Phys. 191, 325
\bibitem[1995]{KR95}
   Kitchatinov~L.~L., R\"udiger~G.: 1995, A\&A 299, 446
\bibitem[1999]{KR99}
   Kitchatinov~L.~L., R\"udiger~G.: 1999, A\&A 344, 911
\bibitem[2000]{KJD00}
   Kitchatinov~L.~L., Jardine~M., Donati~J.-F.: 2000 MNRAS 318, 1171
\bibitem[1994]{KPR94}
   Kitchatinov~L.~L., Pipin~V.~V., R\"udiger~G.: 1994, Astron. Nachr. 315, 157
\bibitem[2004]{Kea04}
   Kov{\' a}ri~Z., Strassmeier~K.~G., Granzer~T. et al.: 2004, A\&A 417, 1047
\bibitem[1980]{KR80}
   Krause~F., R\"adler~K.-H.: 1980, Mean-field magnetohydrodynamics and dynamo theory.
   Akademieverlag, Berlin
\bibitem[2004]{PDC04}
   Petit~P., Donati~J.-F., Collier Cameron~A.: 2004, Astron. Nachr. 325, 221
\bibitem[2004]{Pea04}
   Petit~P., Donati~J.-F., Oliveira~J.~M. et al.: 2004, MNRAS 351, 826
\bibitem[1989]{R89}
   R\"udiger~G.: 1989, Differential rotation and stellar convection: sun and
   solar-type stars. Gordon \& Breach, New York
\bibitem[2004]{REKK04}
   R\"udiger~G., Egorov~P., Kitchatinov~L.~L., K\"uker~M.: 2004, A\&A, submitted.
\bibitem[2004]{RH04}
   R\"udiger~G., Hollerbach~R.: 2004, The magnetic Universe. Weinheim: Willey.
\bibitem[1998]{Sea98}
   Schou~J., Antia~H.~M., Basu~S. et al.: 1998, ApJ 505, 390
\bibitem[2003]{S03}
   Strassmeier~K.~G.: 2003, The solar-stellar connection ... and disconnection.
   In IAU Symposium 219. Stars as Suns: Activity, Evolution and Planets, p.39.
\bibitem[2003]{SKW03}
   Strassmeier~K.~G., Kratzwald~L., Weber~M.: 2003, A\&A 408, 1103
\end{thebibliography}
\end{document}